\documentstyle[epsfig,preprint,aps,prl,floats,epsf]{revtex}
\def\BibTeX{{\rm B\kern-.05em{\sc i\kern-.025em b}\kern-.08em 
    T\kern-.1667em\lower.7ex\hbox{E}\kern-.125emX}}
\draft 

\begin{document}
\author{F.T. Arecchi$^{*}$, R. Meucci, E. Allaria, A. Di Garbo$^{\dagger }$ and L.S. Tsimring$^{+}$}
\title{Delayed Self-Synchronization in Homoclinic Chaos}
\address{Istituto Nazionale di Ottica Applicata,\\
Largo E. Fermi 6, 50125 Florence\\
Italy.\\
$^{*}$Also the Departement of Physics, University Of Firenze,
Italy\\
$^{\dagger }$Also the Istituto di Biofisica CNR, Pisa, Italy\\
$^{+ }$Institute for Nonlinear Science, University of California, San Diego, La Jolla , CA 92093-0402
}

\date{\today }
\maketitle

\begin{abstract}
The chaotic spike train of a homoclinic dynamical system is self-synchronized by re-inserting 
a small fraction of the delayed output. Due to the sensitive nature of the homoclinic chaos 
to external perturbations, stabilization of very long periodic orbits is possible. 
On these orbits, the dynamics appears chaotic over a finite time, but then it repeats with 
a recurrence time that is slightly longer than the delay time. The effect, called delayed 
self-synchronization (DSS), displays analogies with neurodynamic events which occur in the 
build-up of long term memories.
\hfill\\

PACS\ numbers :  05.45.Xt, 05.45.Vx, 42.65.Sf.

\newpage\ 

\end{abstract}

The last decade has seen a great interest in controlling chaos using small perturbations
\cite{uno}.
It started with the seminal paper by Grebogi, Ott, and Yorke \cite{due} in which they proposed
a method of stabilizing unstable periodic orbits using tiny, yet carefully chosen perturbations
to an accessible system parameter. Subsequently, other methods were proposed, as the delayed 
feedback due to Pyragas \cite{tre}, who showed that re-inserting a time-delayed version of 
the output back into the system can at certain conditions stabilize some of 
its periodic orbits.
However, that method does not provide a reliable procedure that guarantees the 
stability of a chosen orbit, in particular it is exceedingly difficult to stabilize 
long periodic orbits. 
An improved version of the delayed feedback was 
represented by the adaptive control \cite {quattro}, experimentally implemented via a Taylor 
expansion which consists in feeding a delayed fraction of the output as well as 
its variation rate \cite{cinque}. 

In this paper we show that for a chaotic system displaying a continuous return to a saddle focus (Shilnikov
chaos \cite{shil}) and a high sensitivity to external perturbations applied near that focus, 
a long-delayed feedback can indeed be used to stabilize very complex sequences of pulses.
We demonstrate this in numerical simulations and experiments with a CO2 laser as well as 
on a return map model of a chaotic pulse generator. However the validity of this scenario 
is much broader, including possible neurodynamic implications. 
We call this effect "delayed self-synchronization" (DSS), insofar as the time signal appears 
as a train of N erratically distributed spikes (N being the ratio between the delay time and 
the average inter-spike interval) which repeat themselves with the same interspike intervals.
The DSS phenomenon should by no means be confused with the synchronization of two distinct
systems \cite{sinchro}. 

Standard control methods are of limited effectiveness for long times insofar as their 
complexity increases with the length of the periodic orbit, as shown e.g. by the Hunt's 
occasional proportional feedback applied to long period orbits \cite{hunt}.
On the contrary, in DSS each return to the saddle focus permits an independent control 
of the corresponding inter-spike interval, and such an operation can be updated as long 
as one wishes, without any increase in complexity. In our laboratory implementation (Fig.1),
a single-mode CO$_2$ laser with intensity feedback is tuned to the parameter range yielding 
homoclinic chaos \cite{alla}.
The intensity output of the laser consists of a sequence of almost identical spikes,
repeating at erratic times because of the homoclinic scrambling, with an average period of 
500 $\mu s. $ (Fig. 2a). Through a delay unit described elsewhere \cite{nove},
a small delayed perturbation (a few percent of the intensity output) is added to the feedback 
signal responsible for homoclinic chaos. As figures 2-b,c show, this small feedback creates 
a sequence of spikes, which is periodic with a period $T_r$ slightly larger than the delay 
time $T_d$ of the feedback loop (DSS). In fact, the DSS adjusts in such a way that the 
spikes of the delayed signal arrive through the delay line in the time interval of 
largest susceptibility, which occurs around $\tau \simeq 150 \mu s.$ before the next large spike
whence $T_r =T_d + \tau$ . The extra time or "refractory time" $\tau$,
which in fact corresponds to the duration of the quenched (zero intensity) time interval 
for each spike, is also measured by the width of the correlation function 
of the free running laser (Fig.3-a).

The autocorrelation function of the DSS signal has large revival peaks separated 
by the peiod $T_r$ (Fig. 3-b).
Figure 4 shows that the minimal signal $\epsilon$ necessary for DSS is constant for long 
delays, down to a delay of $\tau \simeq 150 \mu s$. For delays below the intrinsic 
refractory time $\tau$, the DSS threshold increases dramatically.

DSS can be further illustrated by the space-time representation (STR) introduced 
elsewhere \cite{dieci}. In this representation, the time series is split into pieces 
of length $T_r$, which then are stacked together as different "snapshots" of a 1D 
spatiotemporal system. Thus, every line of this representation is mapped onto the next line. 
The STR for the free-running system and for the DSS regime are shown in Fig. 5. 
When DSS is achieved, the STR of the dynamics changes drastically. 

We also investigate DSS by a numerical experiment with the model of the 
homoclinic chaos in CO$_2$ laser, introduced earlier \cite{undici},
\begin{equation}
\begin{array}{l}
\dot{x}_1=k_0x_1(x_2-1-k_1\sin ^2(x_6)) \\ 
\dot{x}_2=-\gamma _1x_2-2k_0x_2x_1+gx_3+x_4+p_0 \\ 
\dot{x}_3=-\gamma _1x_3+x_5+gx_2+p_0 \\ 
\dot{x}_4=-\gamma _2x_4+gx_5+zx_2+zp_0 \\ 
\dot{x}_5=-\gamma _2x_5+zx_3+gx_4+zp_0 \\ 
\dot{x}_6=-\beta (x_6+b_0-rX_1)
\end{array}
\end{equation}
Here $x_1$ is the laser intensity, $x_2$ is the gain, $x_3$ , $x_4$, $x_5$ are variables 
of the gain medium, x6 is the feedback voltage, and $p_0$ is the pump rate. 
The delayed feedback is introduced in the equation for the feedback voltage 
$x_6$ via $X_1=x_1(t)+ \varepsilon \cdot \left( x_1\left( t-T_d\right) -\bar{x}%
_1\right)$, where $\bar{x}_1$  is the mean value of $x_1 (t)$.
Parameters of the model at which the regime of homoclinic chaos is observed 
are $k_0 = 28.57$, 
$k_1 = 4.55$; $\gamma_1 = 10.06$, $\gamma_2= 1.06$, $g = 0.05$, $p_0 = 0.016$,
$z = 10$, $r = 70$, $b_0 = 0.173$, $\beta = 0.42$.

In numerical simulations, we are not limited in the length of the time delay and the time 
series. We ran very long simulations (up to $5 \cdot 10^6$ non-dimensional time units) with 
long time delays up to 8000. The numerical time series (Fig. 6) are very similar to 
the experimental ones. The auto-correlation function also exhibits sharp peaks separated 
by a time interval $T_r$ that is slightly larger than $T_d$. 
The magnitude of the peaks very slowly decays with time,
which indicates the slow evolution of the periodic orbit.

Notice that, once the dynamical system has become periodic with period $T_r = T_d +\tau$,
that is, $x_1(t)=x_1(t-T_r)$, than by Taylor expansion we can write
$X_1=( 1+\epsilon ) \cdot x_1(t)+\epsilon \cdot (
\frac{dx_1}{dt}) \tau -\epsilon \cdot \bar{x}_1$.
Thus for $\varepsilon =-1$, the delay feedback effect disappears as zero-th order term, 
as expected from Pyragas \cite{tre}, but a first order correction proportional 
to the time derivative of $x_1$ represents the adaptive correction of Ref.\cite{quattro}
in the approximation of Ref.\cite{cinque}.

The DSS mechanism can be elucidated using the following simplified model.
Without time-delayed feedback, after a pulse is produced, the phase trajectory spends 
some amount of time (which fluctuates because of the homoclinic chaos) near the homoclinic 
saddle before it leaves its neighborhood and produces the next pulse.
A close inspection of the time series, both experimental and numerical (see Figs. 2-b,c and 6),
indicates that the arrival of a pulse through the delay loop triggers the escape of the 
phase trajectory of the laser system from the neighborhood of the homoclinic saddle.
Thus, the following simplified model can be proposed. Consider a system that generates 
a pulse at time $t_i$ , which is determined by a map 
\begin{equation}
\begin{array}{l}
t_i=t_{i-1}+f\left( t_{i-1}-t_{i-2}\right) 
\end{array}
\end{equation}
If the function $f\left( \bullet \right)$ is chosen such that the map 
$x_{n+1}=f\left( x_n\right)$ produces chaos, then the sequence of pulses will be chaotic.
Such system has actually been implemented in electronic circuitry \cite{dodici}.
Let us assume now that the pulses are inserted back into the system after a certain time 
delay $T_0$, so if a delayed pulse arrives at a time $t_p$ between $t_{i-1}+\delta$ 
and $t_i$, then it triggers a new pulse at $t_p$ instead of $t_i$,
and the system is reset so that new $t_i=t_p$.
We introduce the small offset $\delta$ in order to avoid spurious generation of pulses 
with very small time distance, which would effectively destroy pseudo-chaotic pulsation. 
In fact in the experiment this "refractory time" was $T_r - T_d = 150 \mu s$.
We iterated this system numerically using the logistic map 
$f\left( x\right) =ax\left( 1-x\right) $ with $a=3.8$ and found that the system typically 
settles on a periodic orbit with the period $T_0$ as shown by 
its autocorrelation function (Fig.7). The particular orbit will obviously depend on initial 
conditions, in particular, orbits with periods much smaller than $T_0$, can also 
be stabilized, however they are not typical. Note, in this case the strength of the feedback 
is irrelevant as long as the magnitude of the pulses passed through the delay line 
is sufficient to trigger the next pulse. 
This agrees with the experimental observation that the threshold of synchronization 
is independent of $T_0$ for large enough $T_0$ (Fig. 4).
We also checked the influence of noise on DSS. 
If noise was added to the sequence of the pulses at the levels below the threshold 
for pulse generation, it would not affect DSS. However, if a small white Gaussian 
noise is added to the inter-pulse interval, the revival peaks eventually decay to zero,
albeit very slowly. 

We have thus introduced a conceptual model for the phenomenon of delayed 
self-synchronization which leads to stabilization of very long-periodic orbits in a 
chaotic system. Such long orbits are known as a signature of the so-called pseudo-chaos 
observed in systems with discretized state space \cite{tredici}. In those systems, 
the short-term behavior is indistinguishable from a real chaos, however the orbit returns 
to the initial point after some time and then repeats itself. 
Obviously, in the systems considered here, the state space is continuous, 
so the analogy between pseudo-chaos and DSS cannot be carried too far. 

The ability to synchronize very long and complex periodic orbits is of particular interest 
in relation to the recent discovery of the neuronal mechanism of transformation of 
short-term memories into permanent (long-term) memories via so called synaptic reentry 
reinforcement (SRR) \cite{quattordici}. 
In those experiments, it has been demonstrated that neurons responsible for learning, 
repeatedly "replay" certain spiking patterns in order to establish permanent connections 
(synapses) among neurons. We believe that stabilization of the complex periodic orbits 
in simple chaotic systems presented in this paper can serve as a paradigm for SRR in more
complex neural systems. In our forthcoming work we will consider stabilization of long 
periodic orbits in the Hindmarsh-Rose neural model \cite{quindici}. 
We believe that the homoclinic chaos observed in that system as well as in real biological 
neurons \cite{sedici}, is analogous to the laser chaos described here, and thus is 
should be susceptible to the phenomenon of delayed self-synchronization.

Authors are indebted to S. Boccaletti and N.F.Rulkov for useful discussions. 
F.T.A., R.M. and E.A. acknowledge partial support from the European Contract 
No. HPRN-CT-2000-158. A.D.G. is supported by European Contract No. PSS 1043. 
L.T. wants to thank Istituto Nazionale di Ottica Applicata (Florence) for warm hospitality. 
The work of L.T. was partially supported by the Engineering Research program of the Office 
of Basic Energy Sciences at the U.S. Department of Energy, grants DE-FG03-95ER14516 and 
DE-FG03-96ER14592 and by the UC MEXUS-CONACYT grant.

\newpage\ 

\begin{center}

{\bf Figures}
\end{center}

\begin{figure}[h]
\centerline{\epsfig{figure=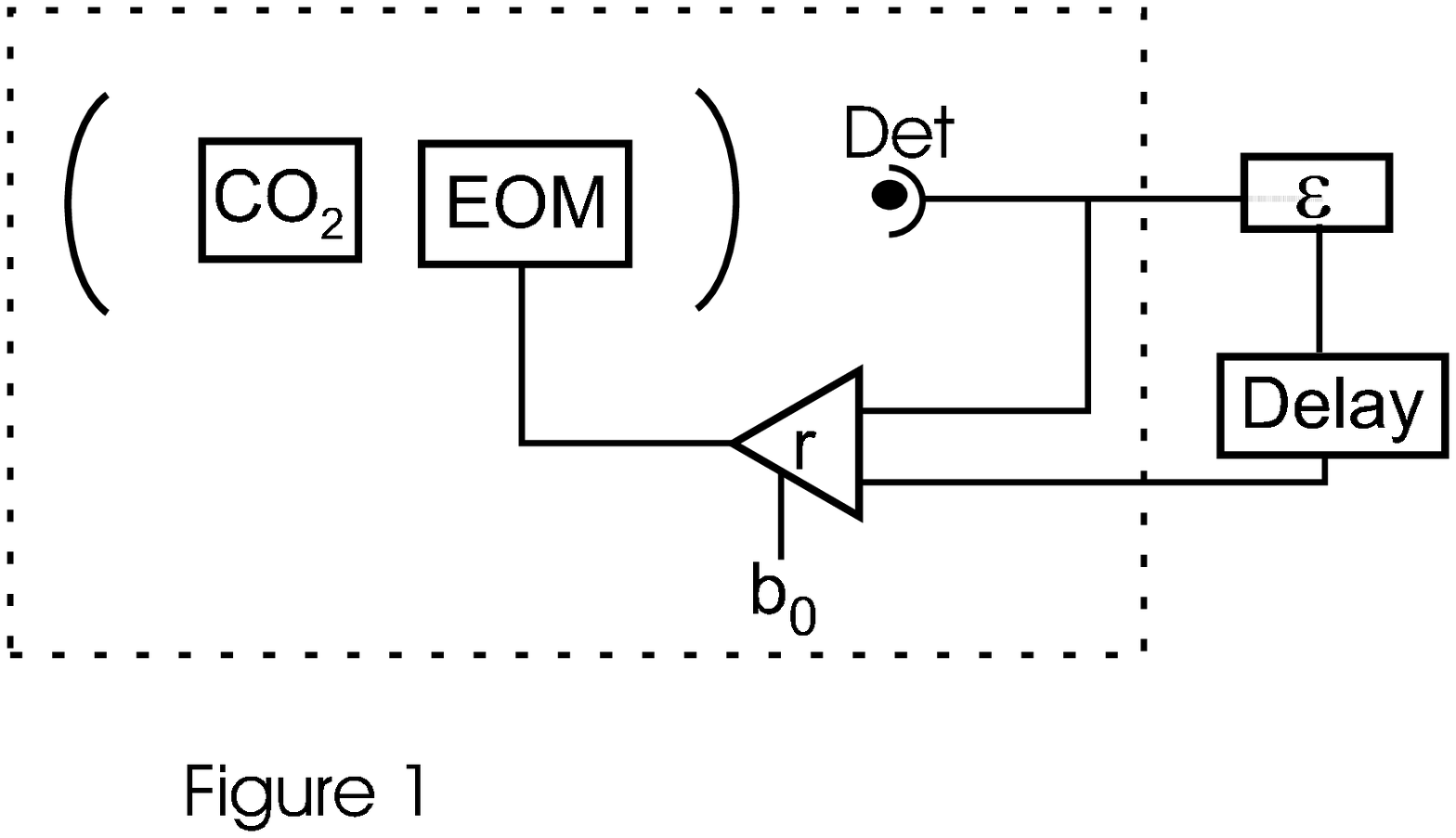,width=7cm}}
\caption{Experimental setup consisting of a CO$_2$ laser with a feedback loop (dashed line)
imposing a regime of homoclinic chaos, and a delay unit. 
$EOM$, electro-optic modulator; $Det$, HgCdTe detector; $r$ and $b_0$, gain and bias of the
amplifier in the feedback loop; $\epsilon$ and $Delay$, coupling factor and delay units.}
\label{figura1}
\end{figure}

\begin{figure}[h]
\centerline{\epsfig{figure=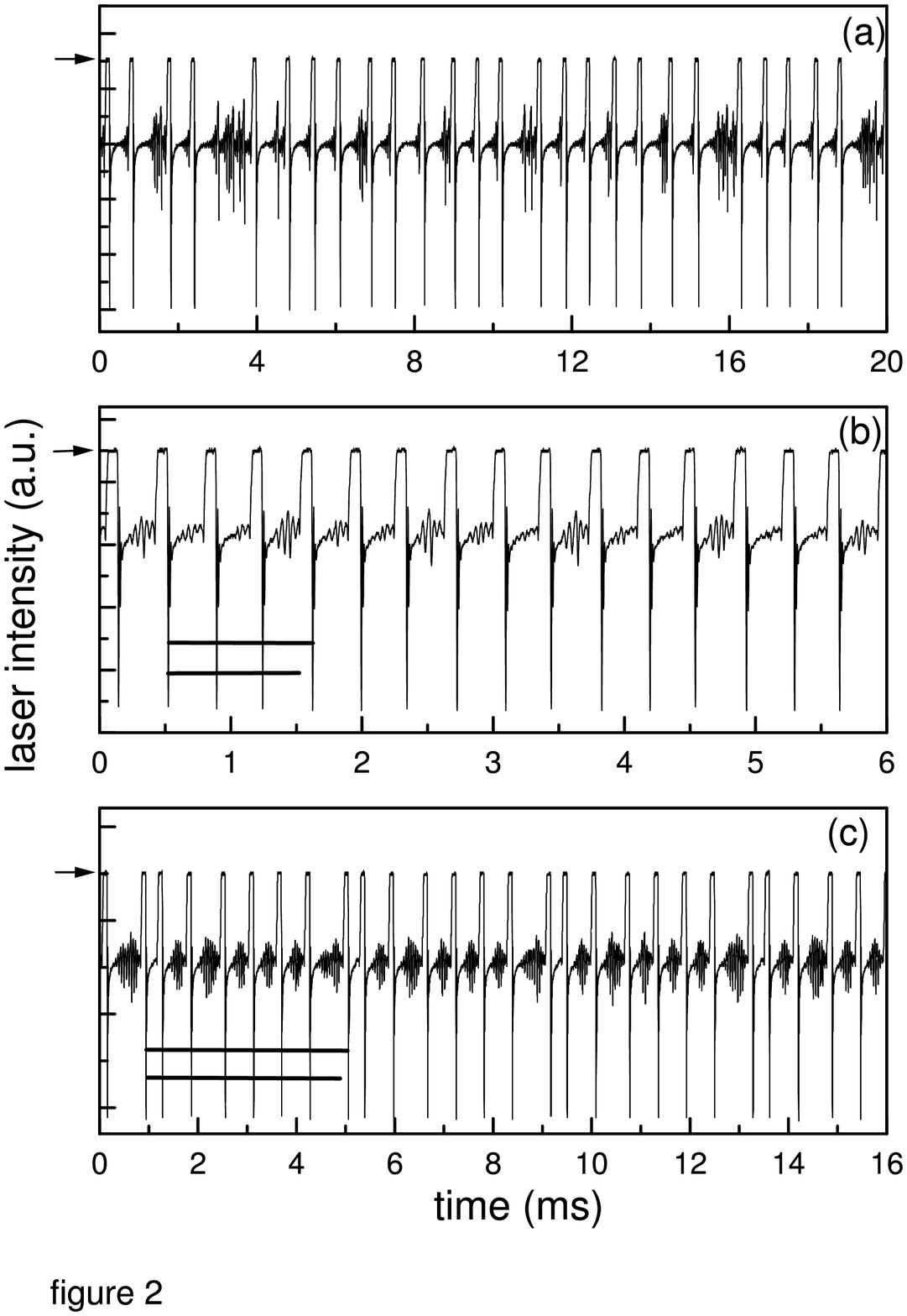,width=7cm}}
\caption{A sequence of homoclinic spikes in the output intensity of a CO$_2$ laser 
in the free running regime (a) and DSS for two different delays (1 and 4 $ms$, (b) and (c)
respectively). A thick arrow denote the zero intensity level. The two horizontal bars for 
1 and 4 $ms$ show the role of the refractory time $\tau=150 \mu s$. The shorter bar is 
the imposed delay ($T_d$), the larger one, with $\tau$ added, is the effective delay 
($T_r$) that characterizes DSS.}
\label{figura2}
\end{figure}

\begin{figure}[h]
\centerline{\epsfig{figure=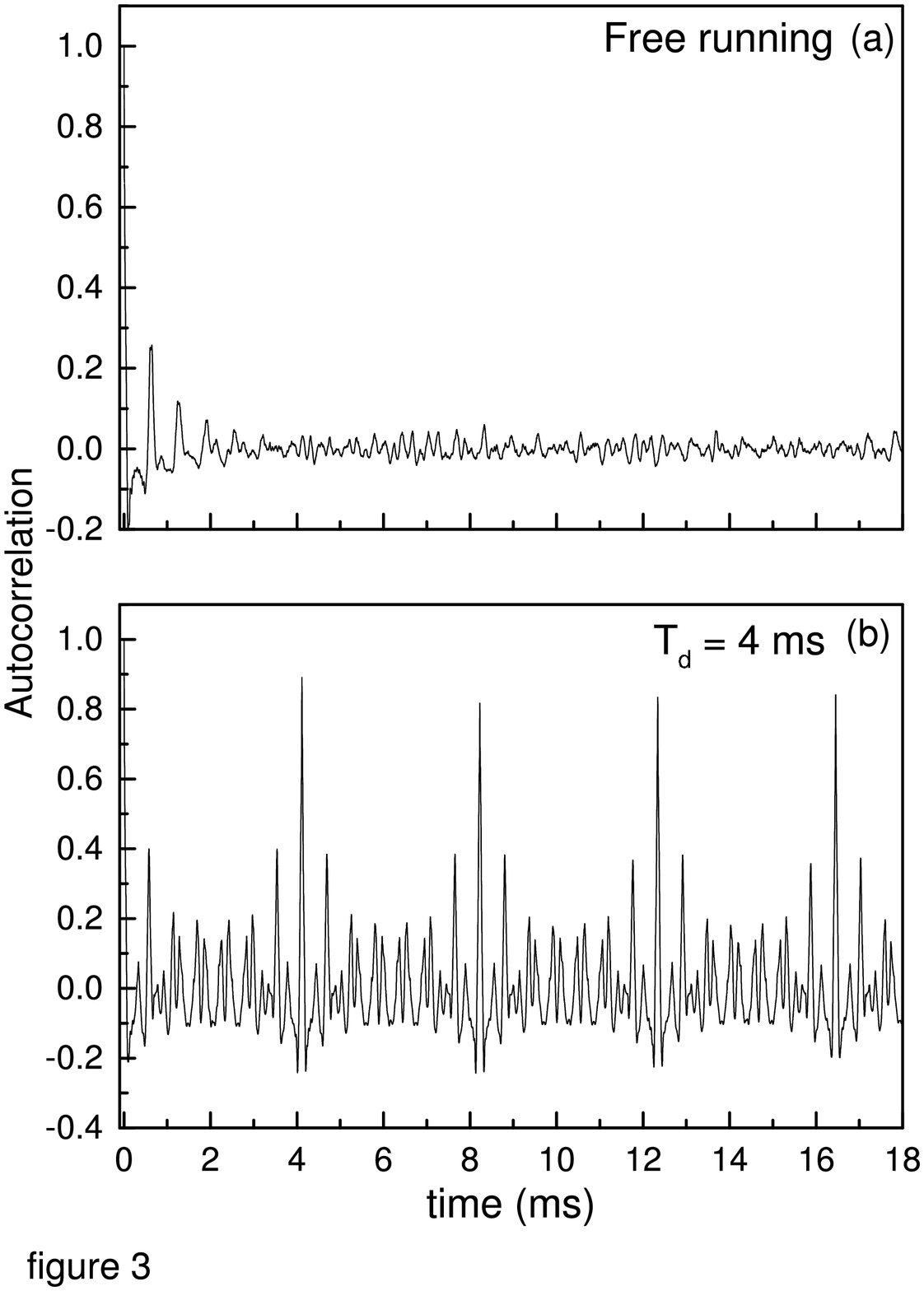,width=7cm}}
\caption{Autocorrelation function of the laser intensity for the free running case (a)
and for DSS with $T_d = 4 ms$ (b). 
The recurrence time $T_r$ for the revival of the correlation is the sum of $T_d$ 
plus the refractory time $\tau$.}
\label{figura3}
\end{figure}

\begin{figure}[h]
\centerline{\epsfig{figure=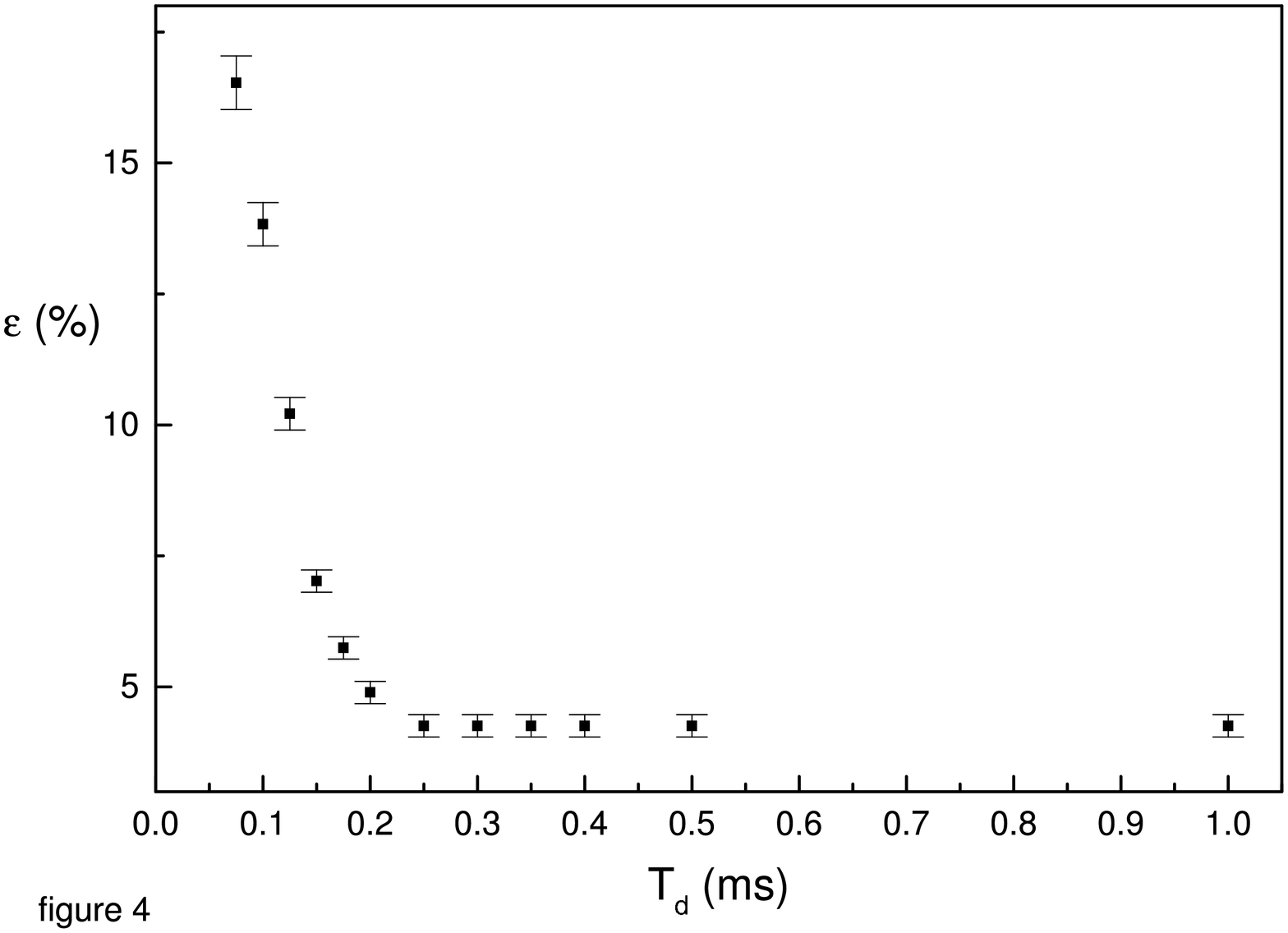,width=7cm,angle=270}}
\caption{Minimal DSS signal (in percent of the peak to peak output) versus the delay time.}
\label{figura5}
\end{figure}

\begin{figure}[h]
\centerline{\epsfig{figure=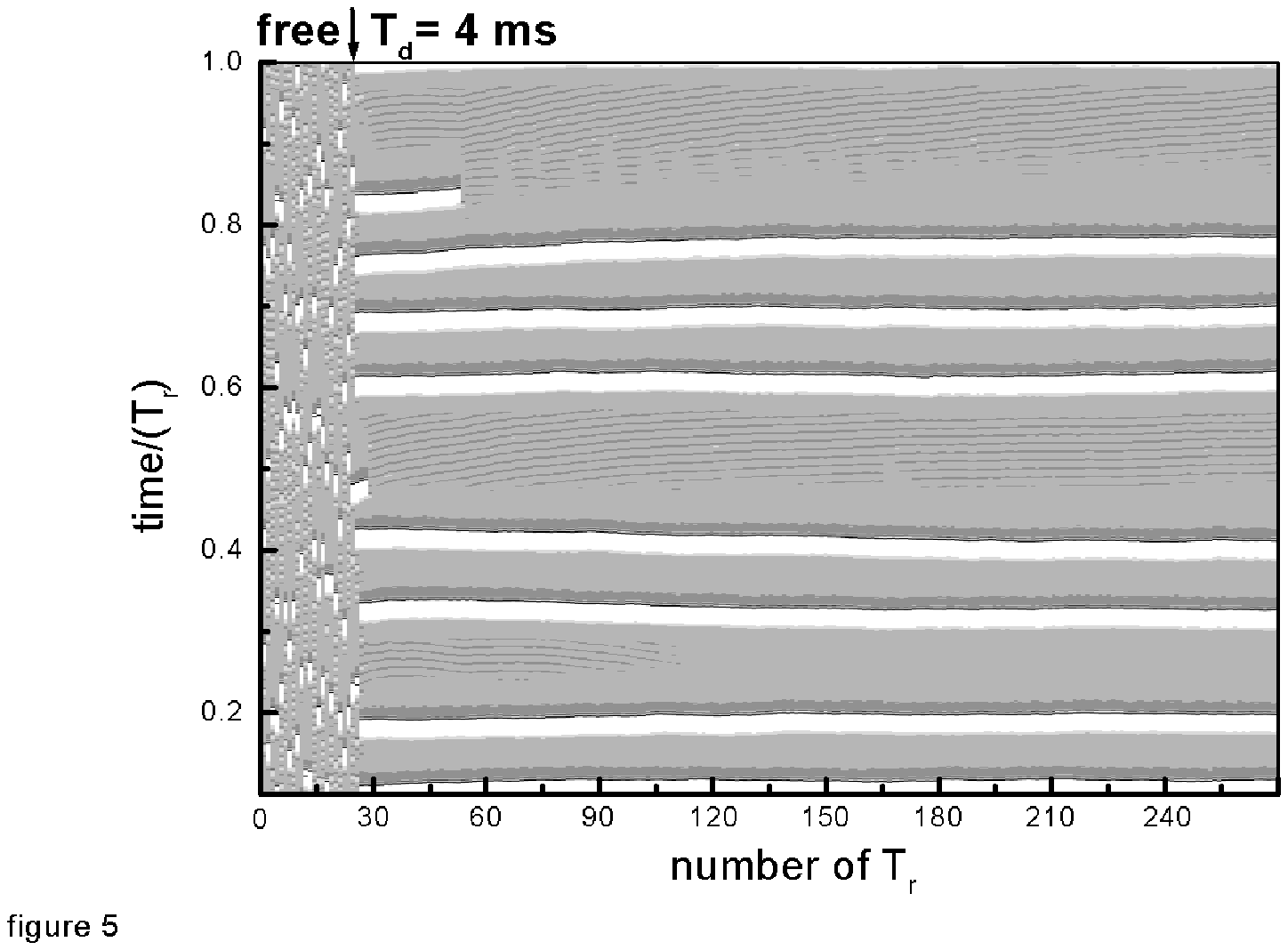,width=12cm,angle=0}}
\caption{Space-time plots of the transition from free running to DSS regime 
with a delay time $T_d = 4 ms$ (indicated by the arrow). The space cell corresponds 
to a single recurrence time $T_r$, the time coordinate is a sequence of integer 
values corresponding to the successive delay units.}
\label{figura6}
\end{figure}

\begin{figure}[h]
\centerline{\epsfig{figure=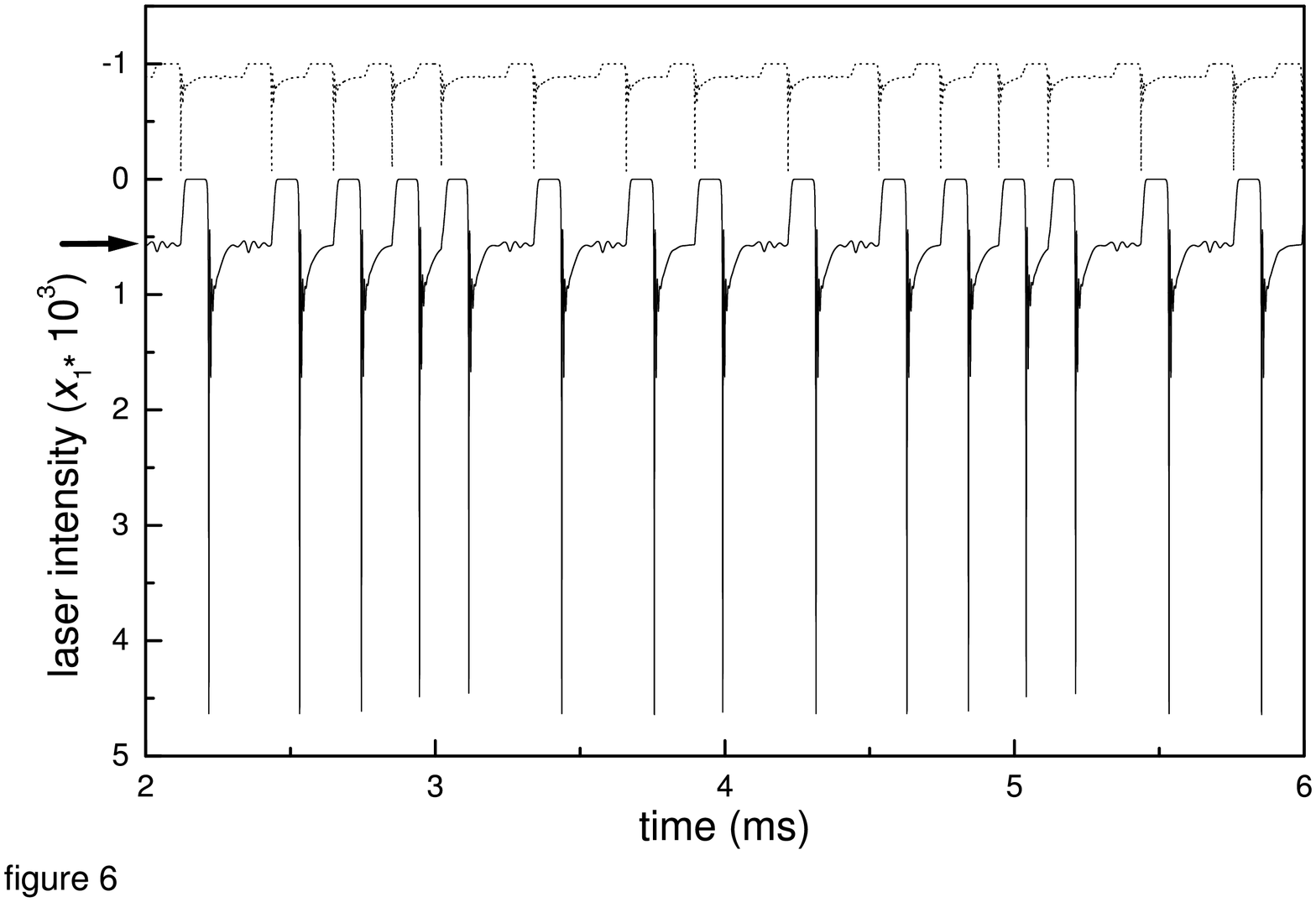,width=7cm,angle=270}}
\caption{Numerical simulations of the laser equations (1) with time delay $T_d=2 ms$: 
solid line, a section of the time series of the laser output;
dotted line - the same time series shifted forward by the delay time $T_d$
and displaced vertically by $-1$. As clearly seen, the spikes of the delayed signal
give rise to the rapid escape of the laser intensity from the floor at $0.5$ 
(denoted by a thich arrow), corresponding to the saddle-focus.}
\label{figura7}
\end{figure}

\begin{figure}[h]
\centerline{\epsfig{figure=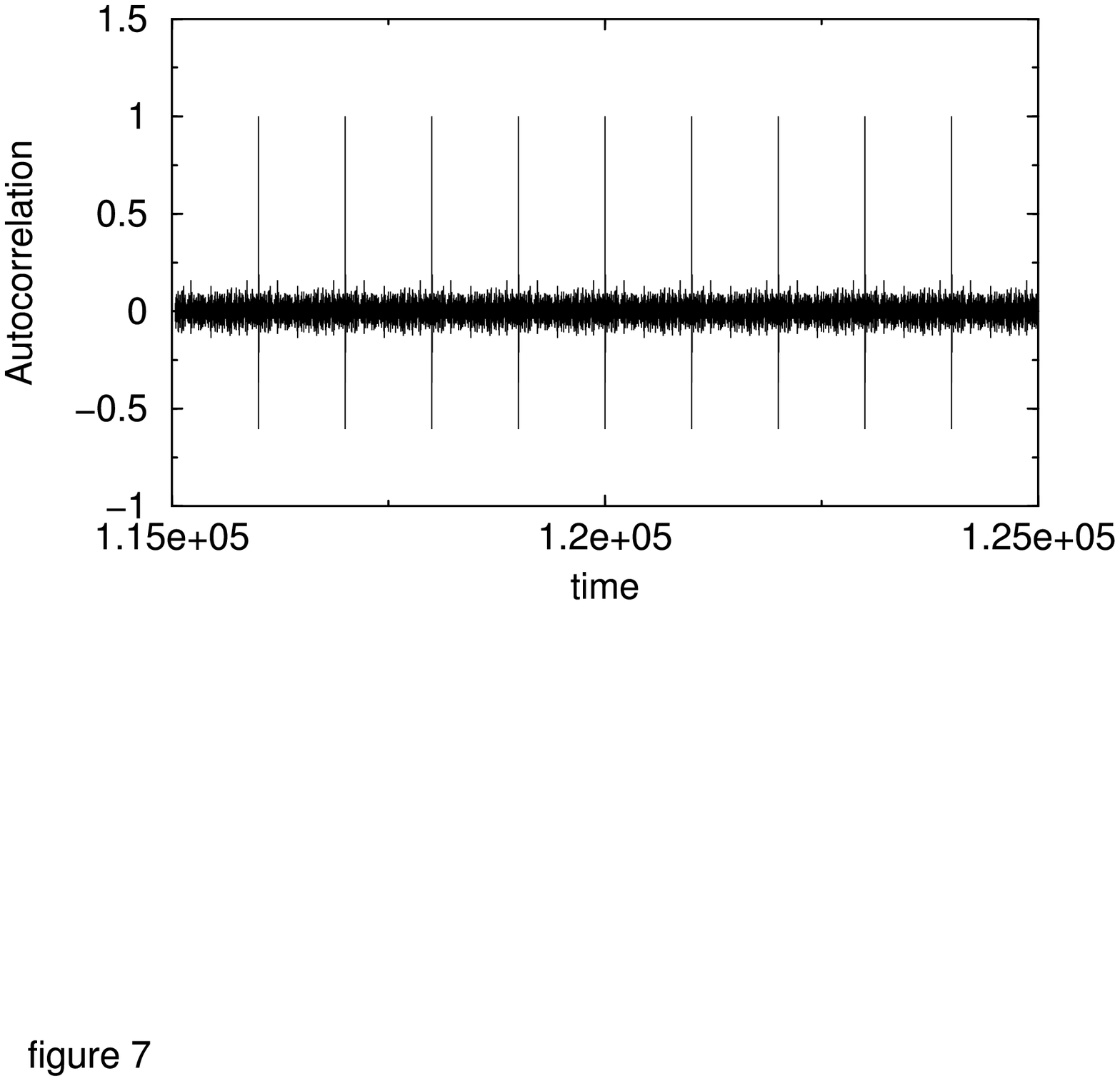,width=9cm}}
\caption{Autocorrelation function of the chaotic pulse generator (2) with delayed feedback. 
Sharp spikes are separated by the delay time $T_0=1000$.}
\label{figura1}
\end{figure}


\begin{thebibliography}{20}
\bibitem{uno}S. Boccaletti, C. Grebogi, Y.-C. Lai, H. Mancini and D. Maza, Physics Report {\bf 329}, 103 (2000).  

\bibitem{due}E.Ott, C.Grebogi, and J.Yorke, Phys. Rev. Lett. {\bf 64}, 1196 (1990).  

\bibitem{tre}K.Pyragas, Phys. Lett, A {\bf 170}, 421 (1992).  

\bibitem{quattro}S. Boccaletti, and F.T. Arecchi,  Europhys. Lett. {\bf 31}, 127 (1995).  

\bibitem{cinque}S. Boccaletti, D. Maza, H. Mancini, R. Genesio and F.T. Arecchi,
Phys. Rev. Lett. {\bf 79}, 5246 (1997);
M. Ciofini, A. Labate, R. Meucci and F.T. Arecchi, Eur. Phys. J. D {\bf 7}, 9 (1999).  

\bibitem{shil}L.P. Shilnikov, Sov. Math. Dokl. {\bf 6}, 163 (1965).  

\bibitem{sinchro} V.S. Afraimovich, N.N Verichev and M.I. Rabinovich, Radiophys. Quant.
Electron. {\bf 29}, 795 (1986);
L.M. Pecora and T.L. Carroll Phys Rev Lett. {\bf 76}, 1916 (1996); 
M.G. Rosenblum, A.S. Pikovsky and J. Kurths, Phys.Rev. Lett. {\bf 78}, 4193 (1997). 

\bibitem{hunt} E.R. Hunt, Phys. rev. Lett. {\bf 67}, 1953 (1991).

\bibitem{alla}E. Allaria, F.T. Arecchi, A. Di Garbo and R. Meucci, Phys. Rev. Lett. 
{\bf 86}, 791 (2001).  

\bibitem{nove}G. Giacomelli, R. Meucci, A. Politi and F.T. Arecchi, Phys. Rev. Lett. 
{\bf 73}, 1099 (1994).  

\bibitem{dieci} F.T. Arecchi, G. Giacomelli, A. Lapucci and R. Meucci, Phys. Rev. A 
{\bf 45}, R4225 (1992).  

\bibitem{undici}A. N. Pisarchik, R. Meucci and F.T. Arecchi ",  Eur. Phys. J. D 
{\bf 13}, 385 (2000).  

\bibitem{dodici}N.F.Rulkov and A.R.Volkovskii, Phys. Lett. A {\bf 179}, 332 (1993).  

\bibitem{tredici}B. Chirikov and F. Vivaldi, Physica D {\bf 129}, 223 (1999).  

\bibitem{quattordici}E. Shimizu, Y-P. Tang, C. Rampon and J.Z. Tsien, Science 
{\bf 290}, 1170 (2000).  

\bibitem{quindici}J. L. Hindmarsh and R. M. Rose, Proc. R. Soc. London, Ser. B 
{\bf 221}, 87 (1984).  

\bibitem{sedici}R.C. Elson, A.I. Selverston, R. Huerta, N.F. Rulkov, M.I. Rabinovich 
and H.D.I. Abarbanel, Phys. Rev. Lett. {\bf 81}, 5692 (1998).  



\end{thebibliography}
\end{document}